# Magnetic field of a linear octupole

E. Arribas[1], I. Escobar[1], R. Ramirez-Vazquez[1], A.C. Marti[2], M. Monteiro[3], C. Stari[2], and A. Belendez[4]

[1]University of Castilla-La Mancha, Applied Physics Department, 02071 Albacete, Spain

[2]Universidad de la República, Montevideo 11400, Uruguay

[3]Universidad ORT Uruguay, Montevideo, Uruguay

[4]Universidad de Alicante, Departamento de Física, Ingeniería de Sistemas y Teoría de la Señal, E-03690, San Vicente del Raspeig, Alicante, Spain

We measure the magnetic field produced by a linear octupole formed by four magnets placed in such a way that both their magnetic moments and their quadrupole moments cancel each other. The magnetic field is measured with the magnetic sensor of a smartphone together with a suitable app, using the Android operating system. The main objective is to determine the dependence of the *x*-component of the magnetic field with distance and compare with the theoretical result that tells us that this dependence is a quintic decrease.

Modern smartphones are equipped with various sensors: accelerometer, gyroscope, proximity sensor, ambient light sensor, capacitive sensors, magnetic sensor, GPS and maybe others[1]. The magnetic sensor allows us to measure the magnetic field and works by making use of the classical Hall effect. This magnetic sensor is being used in physics laboratories generating interesting practices for first year students of STEM degrees[1-6].

**Basic Theory**

We know that a point charge creates an electric field that decreases *quadratically* with distance. A *cubic* decrease is typical for the electric or magnetic field of a dipole (electric or magnetic). For a quadrupole it is *quartic* and for the octupole it is *quintic*. This is what we want to investigate in this experiment.

The magnetic field of an octupole is given by[7]

$$B(x) = \frac{48\mu_0 m a^2}{x^5} \tag{1}$$

where $\mu_0$ is the permitivity of free space, *m* is (the magnitude of the vector) magnetic dipolar moment of the magnets, going from the South to the North pole, and *a* is the width of the magnet, in such a way that the quadrupole is *d*= 4*a*.

In this experiment we use four ferrite magnets with equal factory magnetic moments. We take the magnets in pairs and construct two quadrupoles so that their magnetic moments are opposite and cancel each other. We then place both quadrupoles in such a way that the sum of the four magnetic moments cancel each other. The quadrupole moment of the system of the four magnets is zero, but not the octupole moment. Once the four parallel magnets are placed, the octupole is as shown in Fig. 1.

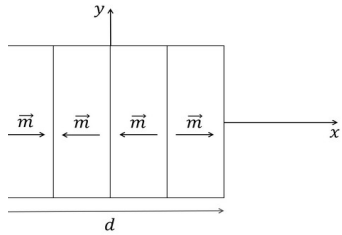

**Figure 1**. Linear octupole formed by four magnetic dipoles. The magnetic moments of each of the four magnets are shown and *d= 4a*, where *a* is the width of a magnet.

**Experimental procedure**

To construct the linear octupole we used four ferrite magnets as shown in Figure 2. The magnets are ring-shaped with 30 mm outer diameter, 23 mm inner diameter and 5 mm width.

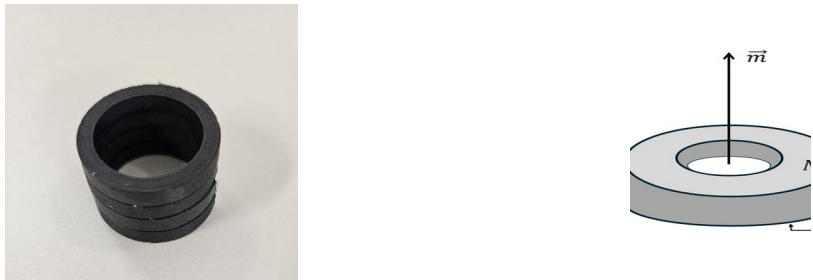

Figure 2. The four ferrite magnets used in this setup. On the right we indicate the magnetic moment of one of them, going from its south pole to its north pole.

We measured the *x*-component of the magnetic field along the *x*-axis using the Physics Toolbox Suite Pro smartphone app with Android operating system[8]. The Pro version is paid ($3) but we recommend it because it allows to calibrate the magnetic field sensor. If the free version is used, it is necessary to measure and subtract the background magnetic field. Before collecting data, it is necessary to find out the precise location of the magnetic sensor of our smartphone, because the *x*-axis must pass through it. We will call *b* the distance to the left of the *y*-axis where it is located. At all times the variable *x* will actually be *x+b* (see Figure 3).

Then we place the octupole at different distances on the *x*-axis and write down the value of the *x* component of the magnetic field measured by the app and the distance *x* (to which we will later add b) between the center of the octupole and the origin of coordinates which is located on the side of the smartphone. The distance must be larger than the size of the octupole if we want to correctly apply Eq. (1).

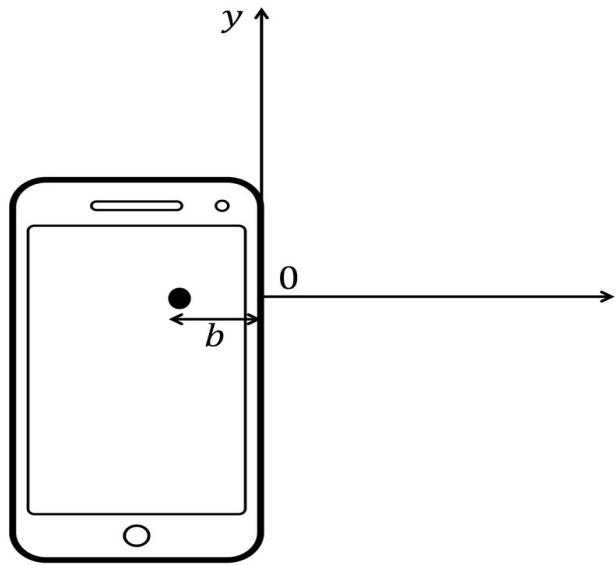

**Figure 3.** Reference system with origin at the edge of the smartphone. The magnetometer is located a distance *b* to the left of the *y*-axis.

**Results**

Figure 4 shows the magnetic field as a function of the distance for the linear octupole. The *x*-axis shows the distance *x+b* in meters and the *y*-axis shows the *x*-component of the magnetic field in µT. Also shown are the potential trend line (power) of these data, its equation and the squared correlation coefficient, whose value is acceptably close to 1, which indicates that the approximation considered to obtain Eq. (1) is fulfilled. As the size of the octupole is 20 mm and the position of the magnetic sensor is $b = 23$ mm, the measurements have been taken from 10.3 cm to 18.3 cm.

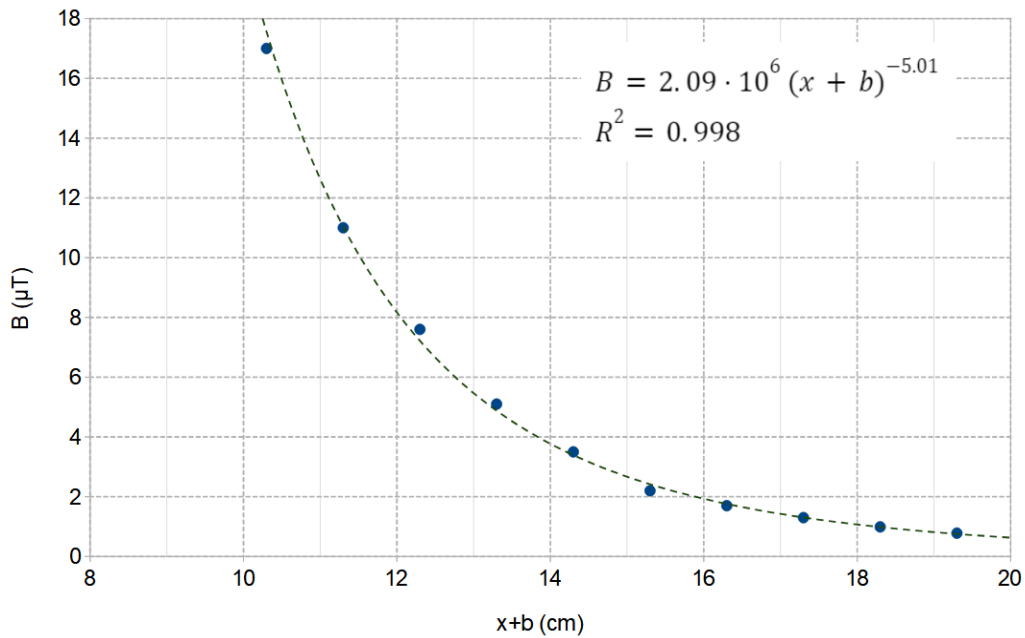

**Figure 4**. Experimental measurements of the *x*-component of the magnetic field along the *x*-axis for a linear octupole with its potential trend line, its equation and the squared correlation coefficient. The number $2.09 \cdot 10^6$ that appears in B's expression has units of $\mu T \cdot cm^5$, it is equivalent to a $2.09 \cdot 10^{-10}$ $T \cdot m^5$.

In Eq. (1) we saw that the dependence of $B$ on distance is $x^{-5}$. To obtain the exponent and its uncertainty from the experimental data we can graph log $B_x$ vs. log $x$ and use the method of least squares fit or linear regression. Modifying Eq. (1) by substituting the exponent by -$n$ and taking logarithms results

$$\log(B(x)) = -n \log(x) + \log(48\,\mu_0 m\,a^2). \qquad (2)$$

By means of the linear regression the exponent is identified with the slope

$$n = 5.01 \pm 0.08 \qquad (3)$$

This experimental value agrees with the expected theoretical value, $n=5$. The relative deviation is 1.6 %.

**Conclusions**

This work can be very easily converted into a laboratory practice for first year students of STEM degrees. Using Eq. (2) the value of the exponent can be obtained by discovery. We are convinced that laboratory work can be very effective in achieving meaningful learning, without the need for the practices to be very sophisticated. They just need to be designed using tools that are familiar to students.

**Acknowledgments**

EA, IE and RRV gratefully acknowledge financial support from the Junta de Comunidades de Castilla-La Mancha of Spain (Project SBPLY/23/180225/00089).

AB gratefully acknowledges financial support from the Generalitat Valenciana of Spain (project PROMETEO/2021/006) and the Ministry of Science, Innovation and Universities of Spain (project PID2021-123124OB-I00).